# Superconducting-like and magnetic transitions in oxygen-implanted diamond-like and amorphous carbon films, and in highly oriented pyrolytic graphite

N Gheorghiu[1*], C R Ebbing[2], J P Murphy[3], B T Pierce[3], and T J Haugan[3]

[1]*UES, Inc., Dayton, OH 45432*

[2]*University of Dayton Research Institute, Dayton, Ohio 45469*

[3]*The Air Force Research Laboratory, Wright-Patterson AFB, Ohio 45433*

*Email: Nadina.Gheorghiu@yahoo.com

**Abstract.** In our previously published work, we have reported colossal magnetoresistance, Andreev oscillations, ferromagnetism, and granular superconductivity in oxygen-implanted carbon fibers, graphite foils, and highly oriented pyrolytic graphite (HOPG). In this follow-up research, more results on these oxygen-implanted graphite samples are presented. We show results from transport measurements on oxygen-implanted diamond-like carbon thin coatings, amorphous carbon films, and HOPG. Significantly, a three-order magnitude drop in the electrical resistance of the oxygen-implanted diamond-like carbon films is observed at the 50 K temperature that we have previously reported for the transition to the superconducting state. Below 50 K, the films' resistance oscillates between the high and low resistance states, less when the sample is under a transverse magnetic field. This metastability between the insulating and superconducting-like states possibly reflects the evolution of the amplitude for the superconducting order parameter also known as the longitudinal Higgs mode. Transitions to low resistance state and metastability are also observed for amorphous carbon films. Finally, the HOPG samples' resistance have a thermally activated term that can be understood on the basis of the Langer–Ambegaokar–McCumber–Halperin model applied to narrow SC channels in which thermal fluctuations can cause phase slips. We also find that in oxygen-implanted carbon materials, the electron charge and spin correlations do not compete and their interplay rather facilitates the emergence of high-temperature superconductivity, and thus, additional unexpected effects like Heisenberg spin waves and magneto-structural transitions are observed.

## 1. Introduction

Following the discovery of high-temperature superconductors/superconductivity (HTSC) like the cuprates, the search for new HTSC materials has been recently directed into ways to avoid the more costly and market restricted rare-earth components or compounds with heavy-metals. One significant, though not always recognized, direction is towards carbon-based or C-content SC materials [1], ranging from several organic, to carbonaceous sulfur hydrides under extremely high pressures [2], and to boron-carbon [3]. The possible role played by sulfur on a-C powders [4] and films [5] has been earlier on investigated.

Carbon (C), hydrogen (H), and oxygen (O) are the building blocks of life. Indeed, the mere existence and evolution of biological structures occurred in the presence of both carbon and water, and life itself is based on hydrogen bonds. In conjugated C-H structures with equal number of C atoms, the delocalized $\pi$ electrons are like the conduction electrons in a metal. Moreover, the $\pi$ electrons are like Cooper pairs in SC and their coherence leads to the Josephson tunnel effect [6]. In these materials, there is competition and interplay between magnetic and SC order [7] and electron viscosity can explain the observation of negative differential resistance and SC [8]. Materials where the building blocks are C and H atoms are practically small SCs, moreover ferromagnetic (FM), as we have previously found [9,10]. On the other hand, it is well-known the important role played by copper-oxide layers in cuprates. For the C-only materials, if instead of powder mixing rather oxygen (O) plasma is used, charge is implanted into the surface of the target film. The observation of room-temperature SC in water-treated graphite [11] suggested that the C can react with oxygen to benefit HTSC.

SC can occur in the bulk or at the surface of a homogeneous system, also at interfaces of a hybrid system. Graphite can have hybrid Bernal (ABAB…) and rhombohedral (ABCABC…) layer stacking. SC was found in the latter [12]. Solution to the tight-binding Bogoliubov-de Gennes equations revealed that for thin stacks (few layers), surface HTSC survives throughout the bulk due to the proximity effect between ABC/ABA interfaces where the order parameter is enhanced [13].

In this paper, we are following up on previously published work [14] by looking for additional signs of SC, possibly even HTSC, in oxygen-implanted C-based thin films as well as graphite bulk samples.

## 2. Experimental

The C-based materials used for this study were diamond-like carbon (DLC) thin films, amorphous carbon (a-C) thin films, highly oriented pyrolytic graphite (HOPG, particle size 10 μm) disc-shaped samples (1 mm thick and 3 mm diameter), and graphite foil [15]. The DLC and a-C films were deposited on sapphire ($Al_2O_3$) substrates using pulsed laser deposition. In order to bring the density of states to levels where possible SC behavior can be observed, the samples were O-ion implanted [16] at two-dimensional (2D) concentrations $7.07 \times 10^{12}$, $5.66 \times 10^{15}$, and $2.24 \times 10^{16}$ ions/cm$^2$, respectively. Details on the ion-implantation efforts can be found in [17], while here we show the data for a-C films and for the graphite foil (Figure 1, left). The implantation energy was 70 keV. For the ion concentrations used, the implantation depth went from ~0.01% to 100% of the either the graphite foil or the a-C samples, from ~0.01% to ~40% for the HOPG samples, and it went all the way though the DLC films (i.e., 100% of the sample). Samples' thickness was: 80 nm for the DLC films, 175 nm for the a-C films, and 1 mm for the graphite foils or the HOPGs. Importantly, the ion implantation changes the hybridization of the C atoms. Thus, as the a-C thin films have a mixture of sp$^2$ and sp$^3$ bonds, the ion implantation changes the sp$^2$/sp$^3$ ratio such that the bandgap of an a-C is tuned between that of diamond and graphite. In a reverse way, the ion implantation of DLC films results in the sp$^3$ to sp$^2$ bond conversion. The four-wire Van Der Pauw square-pattern technique was used for resistivity measurements. The quality of the silver electrical contacts was optically checked using an Olympus BX51 microscope. Raman spectra showed the effect of O-implantation (Figure 1, right). Magneto-transport and magnetization measurements were carried out in the 1.9 K - 300 K temperature range and for magnetic fields of induction *B* up to 9 T using the Quantum Design Physical Properties Measurement System (PPMS) model 6500.

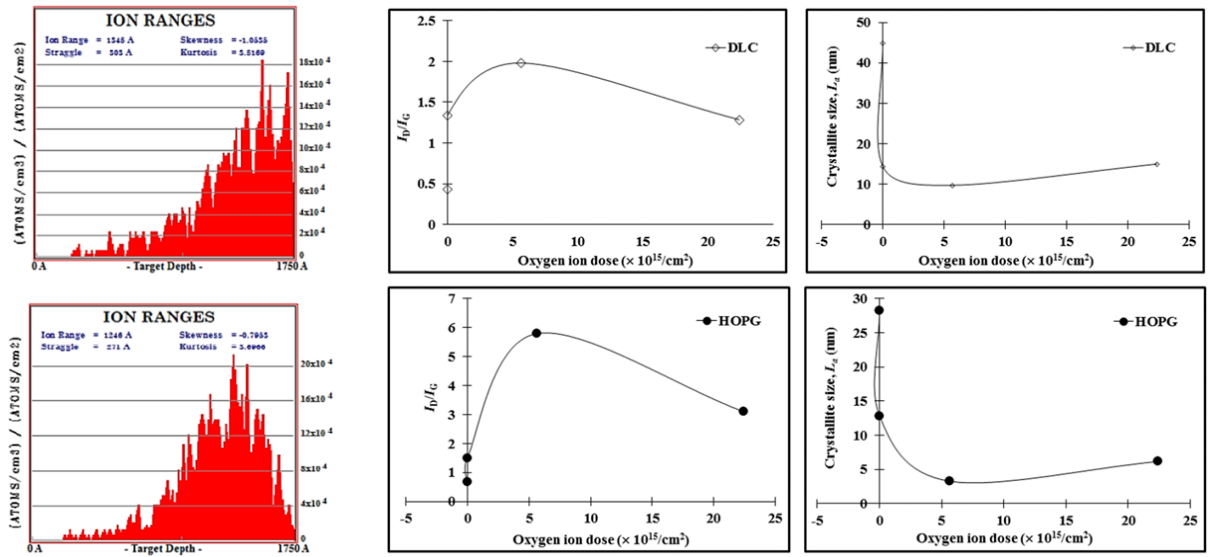

**Figure 1.** Left: 70 keV ion implantation of oxygen in a-C (top) and graphite (bottom), respectively. Right: The effect of O-implantation concentration on the Raman peak $I_D/I_G$ intensity ratio and on the G peak wavevector for the DLC sample (upper plots) and for the HOPG sample (lower plots), respectively.

## 3. Experimental Results
*3.1 Superconducting-like transitions and metastability in oxygen-implanted diamond-like C thin films*

Temperature-dependent measurements of the electrical resistance $R$ for the DLC films revealed a SC-like transition below $T_c \sim 50$ K (Figure 2). $R$ oscillates between $R_{max}$ and $R_{min} \sim R_{max}/1000$. Below $T_c \sim 50$ K, we observe metastability between the SC and the normal state in this physical system. As known, H.K. Onnes' proof of discovering SC in mercury below $T_c \cong 4.19$ K was the ×10,000 decrease in the electrical resistance. The metastability found here for DLC films reflects oscillations in the amplitude SC order parameter otherwise known as the Higgs mode. We also found that a magnetic field applied normal to the DLC film (i.e., along the *c*-axis) appears to weaken and possibly destroy the metastability. The weakening effect might be due to the coexistence between singlet SC and antiferromagnetism (AFM) goes through a first-order phase transition to the coexistence between triplet SC and FM as $B$ strength is increased. This transition was predicted for the occurrence of surface or interface SC in rhombohedral graphite [7]. We have found proof for the AFM-SC (singlet SC) and FM-SC (triplet SC) coexistence below $T_c \sim 50$ K also in hydrogenated C fibers under high magnetic field when comparing the measurements taken during cooling and heating, respectively [9]. On the other hand, the effect played by the magnetic field on cancelling the metastability and lowering of $T_c$ (Figure 2c) relates to the case known for most SC materials where a magnetic field $B \| c$ actually destroys the SC state.

Figure 2 also reveals negative resistance dominating from $T \sim 50$ K to $T \sim 200$ K. Negative $R$ is associated with the viscous flow of electrons, where an electric current running against the applied field (a hole current) is responsible for the observation of a negative nonlocal voltage. The latter may play the same role for the viscous regime as zero electrical resistance does for SC [8]. In this case, possible short-range SC fluctuations starting at $T \sim 200$ K eventually grow into long-range SC fluctuations that become critical at $T_c \sim 50$ K. The oxide layer might host both hard and soft SC fluctuations, i.e., the system might be a multigap SC where the larger gap SC phase takes over the smaller gap SC phases. The evolution of the attractor – which is a manifold of equilibrium states – is actually the evolution of the amplitude (Higgs-Anderson mode) and phase for the SC order parameter.

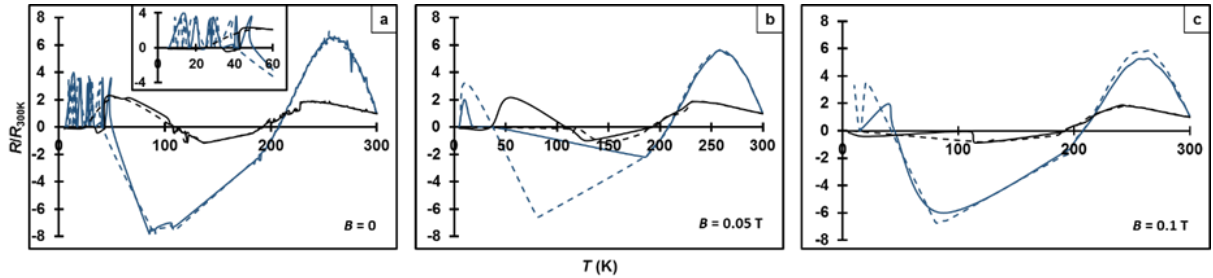

**Figure 2.** Temperature dependent resistance data (relative to the resistance at 300 K) for DLC films that have been O-implanted at 2.24 x $10^{16}$ ions/cm$^2$ (blue) vs. raw samples (black), without (a) and with a transverse magnetic field $B_\perp = 0.05$ T (b) and $B_\perp = 0.1$ T (c), respectively. Dash/full line for data acquired during decreasing/increasing the temperature.

Four fluctuation modes can lead to HTSC in flat-band energy systems like the graphitic ones: the total phase (Nambu-Goldstone) mode, the total amplitude (Higgs-Anderson) mode, the relative phase (Leggett) mode, and the relative amplitude mode [18]. At reasonably high temperatures, the Ginzburg-Landau theory predicts that the amplitude mode contribution dominates because of its singularity at $T = T_c$. At low temperatures, the amplitude mode still has a significant contribution.

We also notice that the insulator-metal transition observed at $T \sim 250$ K is seemingly not affected by $B$. Notice that we have observed the same insulator-metal transition at $T \sim 250$ K also in O-implanted carbon fibers [10]. It is known that fullerene C60 buckeyball clusters could be formed as a result of O-ion implantation. The metal-insulator transition at T ~ 250 K could be a structural phase transition. In the solid state, the C60 molecules assume a simple cubic unit cell that exhibits a phase transition (upon increasing the temperature) to a face-centered-cubic structure upon heating around a transition temperature ~ 250 K [19].

*3.2 Superconducting-like transitions in O-implanted amorphous-C thin films*

Transition to a possible SC state was also found in amorphous-C (a-C) thin films (Figure 3). The conduction goes from electron to hole carrier type (and vice-versa) in more than one occasion and at different temperatures. Notice also that the sample that has been O-implanted at the largest dose goes to superfluidity ($\rho$ becomes very "negative", i.e., the measured voltage is negative) at $T_c \cong 50$ K. Below $T_c$, $\rho$ went beyond the measurable range are thus the apparent lack of data. Figure 3 also shows that significant changes occur in all samples at a temperature around 80 K. One might also question if $T_{BG}$ ~ 80 K is the Bogoliubov-deGennes temperature for this system. For unconventional SC, electron scattering is due to the interaction with the Bogoliubov excitations or bogolons. In this way, bad conductors in the normal phase (like the ones here) can be good SCs for which an alternative mechanism for high-temperature bogolon-pair mediated SC was proposed [20]. The bogolon-pair-mediated scattering can even dominate over the conventional acoustic phonon channel, over the single-bogolon scattering, and over the scattering on impurities. Thus, $T_{BG}$ is larger than $T_c$ for a phonon-based SC. In a hybrid Fermi-Bose system like graphene and possible the one here, $T_c$ can be larger than $T_{BG}$.

Just like was the case with DLC films, metastability in $\rho(T)$ is also observed with these a-C films at temperatures below $T$ ~ 80 K. This was also observed in indium oxide SC films [21]. The accepted caveat is that in disordered 2D systems, SC is only marginally stable and finite low-$T$ resistivity is always expected [22]. The role of disorder in these amorphous films is, indeed, worth mentioning. Disorder-induced granularity was previously found in amorphous indium oxide SC films [23]. Moreover, moderate disorder actually enhances SC, which can persist up to the critical disorder needed to bring about the Anderson transition. In addition, percolation between grains leads to SC, as we have also found [10]. As for the role of non-magnetic impurities and in contrast to the Anderson theorem, note also that a dirty SC does not always mean weak SC [24].

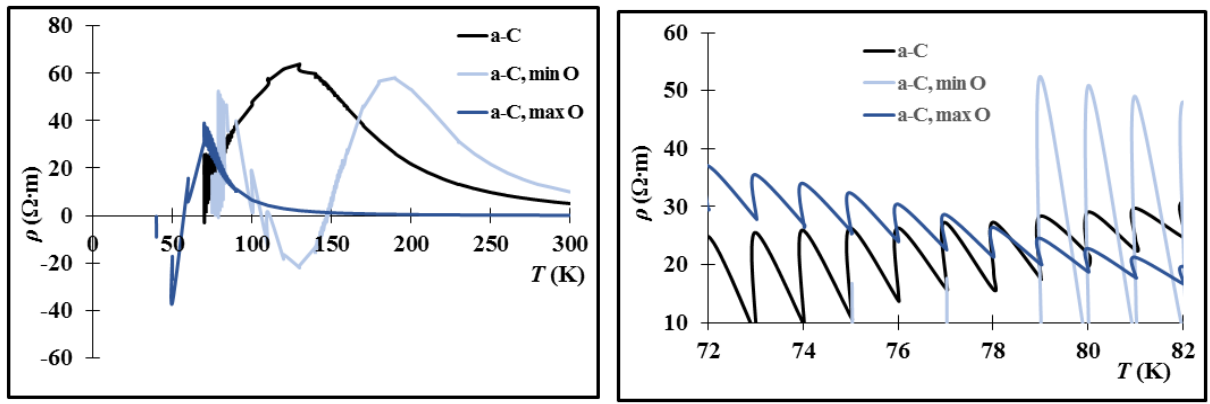

**Figure 3.** Temperature dependent resistivity data at cooling for a-C films: without ion implantation, O-implanted at $7.07 \times 10^{12}$ ions/cm$^2$ (min O), and at $2.24 \times 10^{16}$ ions/cm$^2$ (max O), respectively. The zoomed data (to the right) shows that significant changes occur in all samples at $T$ ~ 80 K, possible the temperature for the exciton condensation (see text).

*3.3 Electric-field driven superconducting-like transition in O-implanted highly HOPG*

While the base carriers in HOPG are electrons, the O-ion implantation results in hole doping of the sample. Temperature-dependent resistivity measurements on the O-implanted HOPG samples revealed the strong dependence on the amount of (direct current) input current $I$. Thus, for $I$ = 20 mA, the data is noisy (Figure 4). While $\rho$ values have the order of magnitude expected for HOPG samples, the highest concentration of O-implantation results in lower values for $\rho$. A larger input current, $I$ = 5 mA, results in smoother data and, more significantly, much lower values for $\rho$ (Figure 5). What is being observed has to do with the expected field effect in this metal-oxide-semiconductor hybrid system, in which the silver contact is the metal, the graphite oxide at the top of the sample is the oxide, and the rest of the sample's bulk – graphite – is the semiconductor. When the strength of the input current is increased, the field effect caused by the edge of the silver contact, results in increasingly deeper in the sample charge injection paths.

Yet, perhaps more significant is the analysis of the cooling-heating cycle for $\rho(T)$ using the two-band model for the electrical conduction [25]. We have modified formula (2) for in [25] for the electrical resistance to include contributions from both electron (*e*) and hole (*h*) charge carriers, in addition, a logarithmic term $\log(T)$ was added. In 2D systems, there are three corrections to the $\log(T)$ term: the weak localization of electrons, the electron-electron interaction for a disordered system, and the Kondo effect. The weak localization and the electron-electron interactions are a favourable brewing ground for SC and an increase in the 2D density of charges through material intercalation can enhance the likelihood for the occurrence of quantum phenomena such as 2D SC [26]. Thus, the total electrical resistance is given by: $R(T) = 1/(R_s^{-1} + R_i^{-1})$, where $R_s = a_1 + b_1\log(T) + a_e(T)\exp[E_{g,e}/(2k_BT)] + a_h\exp[E_{g,h}/(2k_BT)]$ is contribution to $R$ that comes from graphite being a semiconductor, while $R_i = R_0 + r_1T + c_1\log(T) + R_{2,e}\exp[-E_{a,e}/(k_BT)] + R_{2,h}\exp[-E_{a,h}/(k_BT)]$ is the contribution to $R$ coming from the HOPG's interfaces. For sample thickness larger than ~50 nm, the contribution to $R$ coming from the interfaces results in metallic and possible even SC behavior [25]. The physical description of the most relevant terms is as follows. $a_{e,h}(T)$ depends on the charge mobility through the mean free path and on details of the band structure through the effective mass. We used Gnuplot to fit the resistivity data $\rho(T)$ collected at the cooling and heating of the HOPG O-implanted at the highest concentration (2.24 x 10$^{16}$ ions/cm$^2$). We used known formula $\rho = R \times A/l$, where $A$ = 2 mm$^2$ is the cross-sectional area and $l$ = 2 mm is the distance between the current leads. For the energy gap $E_g$ we find: $E_{g,h} \cong 34$ meV and $E_{g,e} \cong -7.8$ meV, respectively. While $E_{g,h}$ is within previously found range, $E_g = (40 \pm 15)$ meV [25], the negative band gap suggests a band overlap (the valence band overlaps slightly the conduction band), i.e., a semimetallic behavior. This was expected, as the thickness of the HOPG sample here is larger than 50 nm. Also obtained from the fit are: the residual resistance $R_0 \cong 56$ μΩ and the coefficient for the linear term $r_1 \cong 0.42 \times 10^{-5}$ Ω/K. Also found are the activation energies for the holes $E_{a,h} = k_B T_{a,h} \cong (1.38 \times 10^{-23}$ J/K)(6.0 K) $\cong 0.52$ meV ($k_B$ is the Boltzmann constant) and for electrons $E_{a,e} = k_B T_{a,e} \cong (1.38 \times 10^{-23}$ J/K)(84 K) $\cong 7.3$ meV, respectively. The activation energies, which are positive, have a special meaning in relation to SC. Namely, the thermally activated term $E_a$ is understood on the basis of the Langer–Ambegaokar–McCumber–Halperin (LAMH) model [27,28] that applies to narrow SC channels in which thermal fluctuations can cause phase slips. For clean HOPG, $T_a \cong 40$ K [25], which is close to the average value here, $T_{a,average} \cong 45$ K. A similar dependence was observed in granular Al–Ge [29], suggesting that granular SC might also exist in the O-implanted HOPG. We have also estimated the thermopower $S_{h,e} = E_{g,e(h)}/(|e|T)$, where $e = -1.6 \times 10^{-19}$ C is the electron's charge. The corresponding values are $S_h \cong 113$ μV/K for holes and $S_e \cong -26$ μV/K for electrons, respectively. Note that $S$ for pristine graphite is nearly zero. By comparison, the thermopower for a classical gas is $S_{class.\ gas} = k_B/|e| \cong 87$ μV/K. While $S$ is negative for YBCO, it can turn positive by increasing the oxygen deficiency.

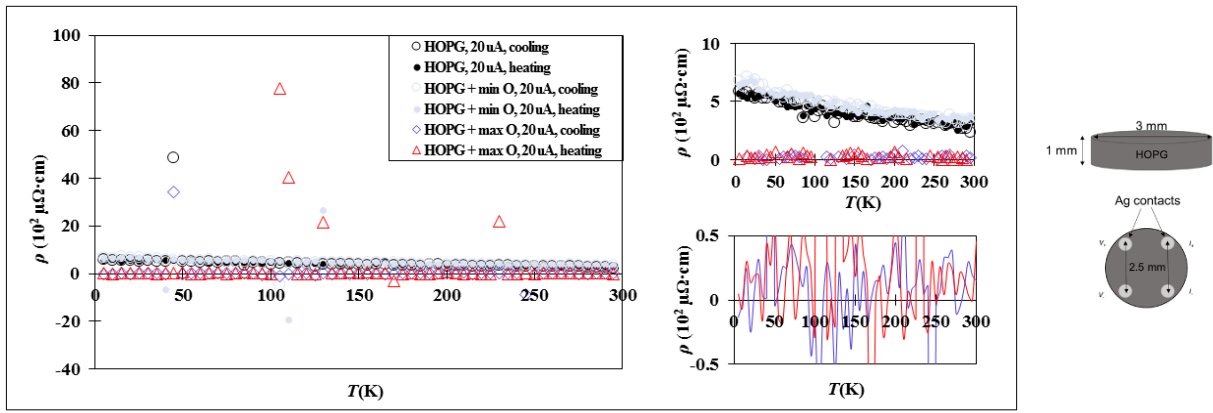

**Figure 4.** Resistivity data for the HOPG samples. O-implantation concentrations: min O = $7.07 \times 10^{12}$ and max O = $2.24 \times 10^{16}$ ions/cm$^2$. The input direct current is small, $I$ = 20 mA. Right: sample and its contacts.

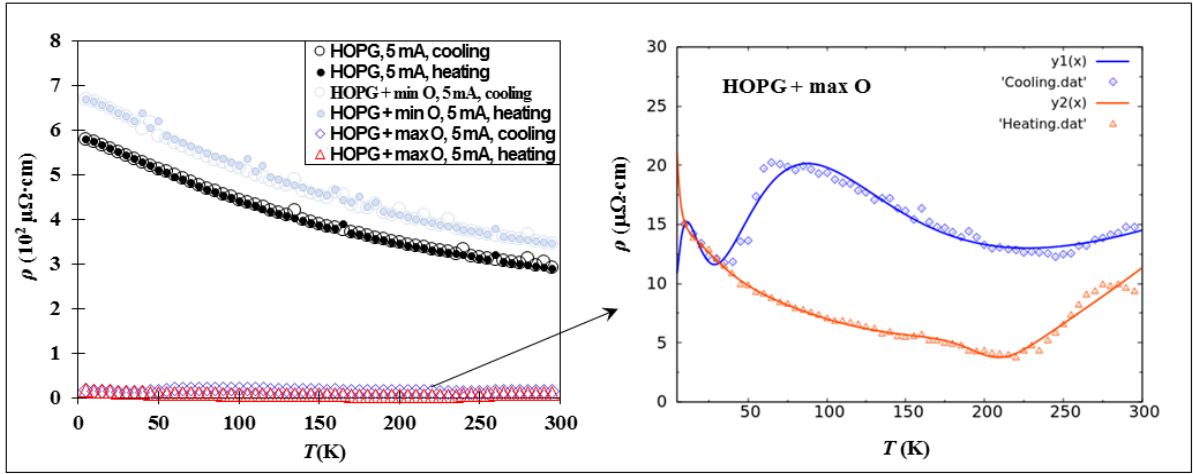

**Figure 5.** Resistivity data for the HOPG samples. Legend for the O-implantation concentrations is the same as in Figure 4. The input current is large, $I = 5$ mA. The fit to the data for the maximum concentration of O-implantation (right) is explained in the text.

*3.4 Other effects in HOPG: Spin waves, magneto-structural transitions, and Almeida-Thouless line*

We have also found that the temperature-dependent remanent magnetization appears to have the behavior consistent with excitations of spin waves in a 2D Heisenberg model with a weak uniaxial anisotropy [30]. Notice that the activation temperature $T_{a,e} \cong 84$ K found before is close to the temperature where the Heisenberg spin wave has an inflection point at zero remanent magnetization. The latter might be related to a magnetic order phase transition from the AFM to the FM order. In addition, there are two magneto-structural transitions, one at $T \cong 80$ K and another, discontinuous, one at $T \cong 210$ K (Figure 6, left). The discontinuity might be due to the fact that some regions still did not complete the transition from AFM to FM. Zero-field cooled and field-cooled temperature-dependent (direct-current induced) magnetization measurements show that the HOPG is at least a spin glass system

(Figure 6, right). Interestingly, $T_c \cong 60$ K (for $H = 0$) extracted from the Almeida-Thouless line is the mean-field $T_c$ for SC correlations in the metallic-H multilayer graphene or in HOPG [31]. The balanced interplay between charge, spin, and lattice orders can favourably result in HTSC.

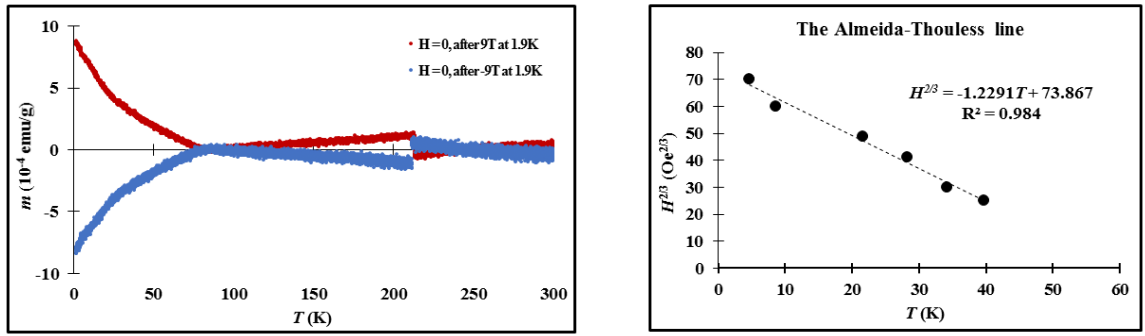

**Figure 6.** Dynamic and static spin behavior in HOPG. Left: Temperature-dependent remanent magnetization. Right: The Almeida-Thouless line.

### 4. Conclusions

Carbon, oxygen, and phase transitions are ubiquitous in nature. In this paper, we bring forth new results from transport measurements on O-implanted DLC coatings, a-C films, and HOPG. Significant drop in the electrical resistance for DLC films is observed at ~50 K, which is the temperature previously reported for the transition to the SC state [9]. The metastability between the normal and the SC state observed below 50 K likely reflects the evolution of the amplitude for the SC order parameter, i.e., the longitudinal Higgs mode. Transitions to low-$R$ states and metastability are also observed in the a-C films. Finally, the HOPG samples' resistance contain a thermally activated term that can be understood on the basis of the LAMH model applied to narrow SC channels in which thermal fluctuations can cause phase slips. Inn O-implanted carbon materials, the electron charge and spin correlations do not compete and together facilitate the emergence of HTSC. Thus, rather unexpected effects like Heisenberg spin waves and magneto-structural transitions are observed.

**Acknowledgments**
This work was supported by The Air Force Office of Scientific Research (AFOSR) for the LRIR #14RQ08COR & LRIR #18RQCOR100 and the Aerospace Systems Directorate (AFRL/RQ). T.J. Bullard is recognized for his initial contribution to this research. N. Gheorghiu acknowledges G.Y. Panasyuk for his continuous support and inspiration.